\documentclass[12pt]{article}

\begin{document}

\begin{center} {\bf Maxwell$-$Chern$-$Simons topologically massive gauge fields in the first-order formalism} \\
\vspace{5mm} S. I. Kruglov\\\vspace{5mm}
\textit{University of Toronto at Scarborough,\\ Physical and Environmental Sciences Department, \\ 1265 Military Trail, Toronto, Ontario, Canada M1C 1A4} \\

\vspace{5mm} \end{center}

\begin{abstract}

We find the canonical and Belinfante energy-momentum tensors and their nonzero traces. We note that the dilatation symmetry is broken and the divergence of the dilatation current is proportional to the topological mass of the gauge field. It was demonstrated that the gauge field possesses the `scale dimensionality' $d=1/2$. Maxwell $-$Chern$-$Simons topologically massive gauge field theory in $2+1$ dimensions is formulated in the first-order formalism. It is shown that $6\times 6$-matrices of the relativistic wave equation obey the Duffin $-$Kemmer$-$Petiau algebra. The Hermitianizing matrix of the relativistic wave equation is given. The projection
operators extracting solutions of field equations for states with definite energy-momentum and spin are obtained. The $5\times 5$-matrix Schr\"{o}dinger form of the equation is derived after the exclusion of non-dynamical components, and the quantum-mechanical Hamiltonian is obtained. Projection operators extracting physical states in the Schr\"{o}dinger picture are found.
\end{abstract}

\section{Introduction}

Physics in $2+1$ dimensions (two spacial dimensions) attracts over past decades
experimentalists and theorists. Thus, the discovery of graphene (planar structure) opens the door to new technology \cite{Geim}. Planar quantum physics differs from the physics in three spatial dimensions. This is due to the existence of the special `Chern$-$Simons' topological term in $2+1$ dimensions (($2+1$)-D). The quantum theory based on the Chern$-$Simons construction leads to fractional quantum statistics and is applied to anyon superconductivity \cite{Wilczek} and the fractional quantum Hall effect \cite{Stone}. Here we consider topologically massive theory in the framework of Maxwell$-$Chern$-$Simons gauge field description \cite{Deser}. Some aspects of Maxwell$-$Chern$-$Simons model were investigated in \cite{Banerjee}, \cite{Horvathy}, \cite{Dunne}, \cite{Colatto}. Chern$-$Simons term violates parity ($P$) and time reversal ($T$) symmetry and is responsible for the mass of the gauge field and spin-1 excitation. Within this theory the gauge invariant mass term possesses the topological nature that is different from the Higgs mechanism of the mass generation. Chern$-$Simons-like topological structures can be considered only in odd-dimensional space-time (for example in $7$-D and $11$-D supergravity). It is valuable that the perturbation quantum theory of the gauge field interaction with fermions in $(2+1$)-D do not have ultraviolet and infrared divergences \cite{Deser} and is super-renormalizable.
Gauge theories in $(2+1$)-D also attract attention because their possible connection with high temperature four-dimensional models.

In this paper, we investigate the behavior of gauge fields under the scale (dilatation) transformations to obtain the
`scale dimensionality' of gauge fields. It is also convenient to consider field equations in the form of the first-order relativistic wave equation (RWE). This allows us to apply well-known covariant methods for obtaining solutions of free gauge field which can be used in different quantum calculations.

The paper is organized as follows. In Sec.2, we obtain the canonical and the symmetrical Belinfante energy-momentum tensors and the dilatation current. The divergence of the dilatation current which is proportional to the topological mass of the gauge field is found. We make a conclusion that the `scale dimensionality' of fields is $d=1/2$.
The field equations are represented in the form of the first-order RWE in Sec.3. We obtain $6\times 6$-matrices of RWE that obey the Duffin$-$Kemmer$-$Petiau (DKP) algebra. Solutions to RWE for free fields with definite energy-momentum and spin are found in the form of projection matrices. The Schr\"{o}dinger form of equations and the quantum-mechanical Hamiltonian are obtained in Sec.4. We find projection operators extracting physical states in the Schr\"{o}dinger picture.  In Sec.5, results are discussed. In Appendix A, the dispersion relation ($p_0^2=\textbf{p}^2 +\mu^2$) is found from the characteristic equation. Appendix B is devoted to obtaining the symmetrical energy-momentum tensor by varying action on metric.

The Euclidean metric is used so that $x_\mu=(x_m,x_3)=(x_m,ix_0)$ and $x_0$ is a time, $x_0=t$, and Greek letters run $1,2,3$ and Latin letters run $1,2$. The system of units $\hbar =c=1$ is explored.

\section{Field equations, the energy-momentum tensor and dilatation current}

We start with the Lagrangian density \cite{Deser}
\begin{equation}
{\cal L} =-\frac{1}{4}F_{\mu\nu}^2+\frac{\mu}{4}\varepsilon_{\mu\nu\alpha}F_{\mu\nu}A_\alpha,
 \label{1}
\end{equation}
where $\varepsilon_{\mu\nu\alpha}$ is antisymmetric tensor, $\varepsilon_{123}=-i$, $F_{\mu\nu}=\partial_\mu A_\nu-
\partial_\nu A_\mu$ is the field strength, and $A_\mu$ is a vector-potential. Thus, there is a coupling of Maxwell and Chern$-$Simons terms in Eq.(1). Note that Chern$-$Simons term is not invariant under parity and time reversal transformations. Equations of motion follow from Lagrangian (1), and are given by
\begin{equation}
\partial_\mu F_{\mu\nu}+\frac{\mu}{2}\varepsilon_{\nu\mu\alpha}F_{\mu\alpha}=0.
 \label{2}
\end{equation}
Eq.(2) is invariant under U(1)-gauge transformations: $A_\mu\rightarrow A_\mu+\partial_\mu \Lambda$.
The dual vector $\widetilde{F}_\mu=(1/2)\varepsilon_{\mu\nu\alpha}F_{\nu\alpha}$ obeys the Klein$-$Gordon equation in $(2+1)$-D with the mass $\mu$ \cite{Deser} (see also \cite{Dunne}). In Appendix A, we demonstrate that Eq.(2) leads to the dispersion relation $p_0^2=\textbf{p}^2 +\mu^2$ which defines the mass $\mu$ of the gauge field $A_\nu$. The dual vector satisfies the equality $\partial_\mu \widetilde{F}_\mu=0$ leaving only two degrees of the vector $\widetilde{F}_\mu$. The gauge field is massive possessing the topological mass $\mu$ and spin-1 \cite{Deser}. The conserved canonical energy-momentum tensor $T_{\mu\nu}^{c}=(\partial_\nu
A_\alpha)\partial{\cal L}/\partial(\partial_\mu
A_\alpha)-\delta_{\mu\nu}{\cal L}$ reads
\begin{equation}
T_{\mu\nu}^{c}=\frac{\mu}{2}\varepsilon_{\mu\alpha\beta}A_\beta\partial_\nu A_\alpha-F_{\mu\alpha}\partial_\nu A_\alpha-\delta_{\mu\nu}{\cal L}.
\label{3}
\end{equation}
One can check using equations of motion (2), that the energy-momentum
tensor (3) is conserved tensor, $\partial_\mu T^c_{\mu\nu}=0$.
The canonical energy-momentum tensor (3) is not the
symmetric tensor, and its trace is given by
\begin{equation}
T_{\mu\mu}^{c}=-\frac{\mu}{2}\varepsilon_{\mu\nu\alpha}F_{\mu\nu}A_\alpha+\frac{1}{4}
F_{\mu\nu}^2. \label{4}
\end{equation}
We will investigate the dilatation symmetry with the help of the method
\cite{Coleman}. Thus, the canonical dilatation current is
\begin{equation}
D_\mu^c=x_\alpha T_{\mu\alpha}^{c}+\Pi_{\mu\nu} d A_\nu,
\label{5}
\end{equation}
where we took into account that the matrix $\textbf{d}$ defining the field dimension is proportional to the unit matrix.
The parameter $d$ is the scale dimensionality of the gauge field $A_\mu$ in $(2+1)$-D. Bose fields in 4-dimensional space-time have $d=1$ and Fermi fields possess $d=3/2$. Calculating
\begin{equation}
\Pi_{\mu\nu}=\frac{{\partial\cal L}}{\partial\left(\partial_\mu A_\nu \right)} =-F_{\mu\nu}+\frac{\mu}{2}\varepsilon_{\mu\nu\alpha}A_\alpha,
\label{6}
\end{equation}
we obtain from Eq.(5),(2)
\begin{equation}
\partial_\mu D_\mu^c=\frac{(d-1)\mu}{2}\varepsilon_{\mu\nu\alpha}F_{\mu\nu}A_\alpha+\frac{1-2d}{4}
F_{\mu\nu}^2.
\label{7}
\end{equation}
It is obvious that for Maxwell theory when $\mu=0$, we have the scale invariance, and the divergence of the dilatation current (7) should vanish. Therefore, we make a conclusion that the scale dimensionality $d$ of the gauge field $A_\mu$ in $(2+1)-D$ is
\begin{equation}
d=\frac{1}{2}.
\label{8}
\end{equation}
Then from Eq.(7), taking into consideration Eq.(8), we obtain
\begin{equation}
\partial_\mu D_\mu^c=-\frac{\mu}{4}\varepsilon_{\mu\nu\alpha}F_{\mu\nu}A_\alpha.
\label{9}
\end{equation}
The divergence of the dilatation current (90) is proportional to the mass parameter as it should be.
One can obtain Eq.(9) also from \cite{Coleman}
\begin{equation}
\partial_\mu D_\mu^c=\Pi_{\mu\nu}(d+1)\partial_\mu A_\nu
 +\frac{\partial{\cal L}}{\partial A_\mu}dA_\mu
 -3{\cal L}.
\label{10}
\end{equation}
The coefficient $3$ instead of $4$ in \cite{Coleman} is connected with the space-time dimension $(2+1)$ .

The non-symmetric and not gauge-invariant tensor (3) can be symmetrized
with the help of the Belinfante trick \cite{Coleman}:
\begin{equation}
T_{\mu\nu}^{B}=T_{\mu\nu}^{c}+\partial_\beta X_{\beta\mu\nu},
\label{11}
\end{equation}
where
\begin{equation}
X_{\beta\mu\nu}=\frac{1}{2}\left[\Pi_{\beta\sigma}\left(\Sigma_{\mu\nu}\right)_{\sigma\rho}
-\Pi_{\mu\sigma}\left(\Sigma_{\beta\nu}\right)_{\sigma\rho}-
\Pi_{\nu\sigma}\left(\Sigma_{\beta\mu}\right)_{\sigma\rho}\right]A_\rho.
\label{12}
\end{equation}
The symmetric Belinfante tensor (11) is also conserved, $\partial_\mu T^B_{\mu\nu}=0$ because the tensor $X_{\beta\mu\nu}$ is antisymmetric in indexes $\beta$ and $\mu$, and $\partial_\mu\partial_\beta
X_{\beta\mu\nu}=0$. The generators of the Lorentz transformations $\Sigma_{\mu\alpha}$ in Euclidian space-time have the matrix elements as follows:
\begin{equation}
\left(\Sigma_{\mu\alpha}\right)_{\sigma\rho}=\delta_{\mu\sigma}\delta_{\alpha\rho}
-\delta_{\alpha\sigma}\delta_{\mu\rho}. \label{13}
\end{equation}
Using equation (13), we find
\begin{equation}
X_{\beta\mu\nu}=\Pi_{\beta\mu}A_\nu.
\label{14}
\end{equation}
With the help of equations (14),(12), the gauge-invariant Belinfante tensor (11) becomes
\begin{equation}
T_{\mu\nu}^{B}=-F_{\mu\alpha}F_{\nu\alpha}+\frac{\mu}{4}\varepsilon_{\mu\alpha\beta}\left(
2F_{\nu\alpha}A_\beta+F_{\alpha\beta}A_\nu\right)-\delta_{\mu\nu}{\cal L}.
\label{15}
\end{equation}
Because the symmetrical energy-momentum tensor for pure Chern$-$Simons field equals zero \cite{Deser} (see also Appendix B), we obtain
\[
T_{\mu\nu}^{B}=-F_{\mu\alpha}F_{\nu\alpha}+\frac{1}{4}\delta_{\mu\nu}F^2_{\alpha\beta}.
\]
Thus, the Belinfante energy-momentum tensor coincides with Maxwell's energy-momentum tensor, and its trace is given by
\begin{equation}
T_{\mu\mu}^{B}=-\frac{1}{4}F_{\mu\nu}^2.
\label{16}
\end{equation}
It is interesting that in (3+1)-D the trace of Maxwell's energy-momentum tensor vanishes in contrast to (2+1)-D.
We define the modified dilatation current according to \cite{Coleman}:
\begin{equation}
D_{\mu}^{B}=x_\alpha T_{\mu\alpha}^{B}+V_\mu,
\label{17}
\end{equation}
where the field-virial $V_\mu$ is
\begin{equation}
V_\mu=\Pi_{\alpha\beta}\left[d\delta_{\alpha\mu}\delta_{\beta\rho}
-\left(\Sigma_{\alpha\mu}\right)_{\beta\rho}\right]A_\rho=\frac{1}{2}\Pi_{\alpha\mu}A_\alpha.
\label{18}
\end{equation}
We obtain from (18),(6)
\begin{equation}
\partial_\mu V_{\mu}=\frac{1}{4}F_{\mu\nu}^2-\frac{\mu}{4}\varepsilon_{\mu\alpha\beta}F_{\mu\alpha}A_\beta.
\label{19}
\end{equation}
As a result, it follows from Eq.(16)-(19) that the equality
\begin{equation}
\partial_\mu D_{\mu}^{B}=\partial_\mu D_{\mu}^c=-\frac{\mu}{4}\varepsilon_{\mu\nu\alpha}F_{\mu\nu}A_\alpha
\label{20}
\end{equation}
holds. The scale (dilatation ) symmetry is broken because the gauge field $A_\mu$ has the topological mass $\mu$.
The conformal symmetry is also broken as the field-virial $V_\mu$ is not the total divergence
of some local quantity $\sigma_{\alpha\beta}$ \cite{Coleman}. The same situation is realized in field theories
where the mass parameter presents (see, for example, \cite{Kruglov}). It should be noted that the scale invariance of Maxwell's field holds in (2+1)-D although the trace is not zero, Eq.(16). It was paid attention in \cite{Banerjee} on the importance of consideration of both symmetrical and canonical (Noether) angular momenta for derivation of the Chern$-$Simons vortices spin.

\section{Relativistic wave equation}

Let us introduce $6$-component wave function
\begin{equation}
\Psi (x)=\left\{ \psi _B(x)\right\} =\left(
\begin{array}{c}
\mu A_\nu(x)\\
F_{\mu\nu}(x)
\end{array}\right),
\label{21}
\end{equation}
where index $B=\nu,[\mu\nu]$, and the function $\Psi(x)$ is the direct sum of the vector $\psi_\nu (x)=\mu A_\nu(x)$, and the second rank antisymmetric tensor $\psi_{[\mu\nu]}(x)=F_{\mu\nu}(x)$.
We use the elements of the entire matrix algebra $\varepsilon
^{A,B}$ \cite{Kruglov1} with matrix elements and products
\begin{equation}
\left( \varepsilon ^{M,N}\right) _{AB}=\delta _{MA}\delta _{NB},
\hspace{0.5in}\varepsilon ^{M,A}\varepsilon ^{B,N}=\delta
_{AB}\varepsilon ^{M,N}, \label{22}
\end{equation}
where $A,B,M,N=\mu,[\mu\nu]$,
and generalized Kronecker symbol is given by
\begin{equation}
\delta_{[\mu\nu][\alpha\beta]}=\delta_{\mu\alpha}\delta_{\nu\beta}- \delta_{\mu\beta}\delta_{\nu\alpha}.
\label{23}
\end{equation}
In the $6\times6$-matrix  $\varepsilon ^{M,N}$ only one element is unity where the row $M$ and the
column $N$ cross and other elements are zero.

With the help of (21)-(23) equation (2) and equation for the field
strength $F_{\mu\nu}=\partial_\mu A_\nu-\partial_\nu A_\mu$ may
be cast in the form of the first-order equation
\begin{equation}
\partial _\mu \left(\varepsilon ^{\nu,[\nu\mu]}+ \varepsilon ^{[\nu\mu],\nu}
\right)_{AB}\Psi _B(x)+ \mu\left(\frac{1}{2}\varepsilon ^{[\nu\mu],[\nu\mu]}+ \frac{1}{2}\varepsilon_{\nu\sigma\beta}\varepsilon^{\nu,[\sigma\beta]}\right) _{AB}\Psi_B(x)=0 ,
\label{24}
\end{equation}
and we imply a summation over all repeated indices. Equation (24) can be written in the form of the
first-order RWE:
\begin{equation}
\left( \beta _\mu \partial _\mu +\mu P\right) \Psi (x)=0 , \label{25}
\end{equation}
where we define $6\times6$-dimensional matrices
\begin{equation}
\beta_\mu=\varepsilon ^{\nu,[\nu\mu]}+ \varepsilon
^{[\nu\mu],\nu},~~~
P=\frac{1}{2}\varepsilon ^{[\nu\mu],[\nu\mu]}+
\frac{1}{2}\varepsilon_{\nu\sigma\beta}\varepsilon^{\nu,[\sigma\beta]}.
\label{26}
\end{equation}
The matrix $P$ is the projection operator, $P^2=P$.  It should be noted that it is not the
Hermitian matrix, $P^+\neq P$. The similar equation was derived for higher derivative
Podolsky's electrodynamics in $(3+1)$-D \cite{Kruglov2}. Matrices (26) can be visualized as follows:
\[
\beta_1=\left(
\begin{array}{cccccc}
0&0&0&0&0&0 \\
0&0&0&-1&0&0 \\
0&0&0&0&1&0\\
0&-1&0&0&0&0\\
0&0&1&0&0&0\\
0&0&0&0&0&0
\end{array}\right),~~~~
\beta_2=\left(
\begin{array}{cccccc}
0&0&0&1&0&0 \\
0&0&0&0&0&0 \\
0&0&0&0&0&-1\\
1&0&0&0&0&0\\
0&0&0&0&0&0\\
0&0&-1&0&0&0
\end{array}\right),
\]
\vspace{-7mm}
\begin{equation} \label{27}
\end{equation}
\vspace{-7mm}
\[
\beta_3=\left(
\begin{array}{cccccc}
0&0&0&0&-1&0 \\
0&0&0&0&0&1 \\
0&0&0&0&0&0\\
0&0&0&0&0&0\\
-1&0&0&0&0&0\\
0&1&0&0&0&0
\end{array}\right),~~~~
P=\left(
\begin{array}{cccccc}
0&0&0&0&0&-i \\
0&0&0&0&i&0 \\
0&0&0&-i&0&0\\
0&0&0&1&0&0\\
0&0&0&0&1&0\\
0&0&0&0&0&1
\end{array}\right).
\]
One can verify that the Hermitian matrices $\beta_\mu$ obey the
DKP algebra \cite{Petiau}, \cite{Duffin}, \cite{Kemmer}:
\begin{equation}
\beta _\mu \beta _\nu \beta _\alpha +\beta _\alpha \beta _\nu
\beta _\mu =\delta _{\mu \nu }\beta _\alpha+\delta _{\alpha \nu
}\beta _\mu . \label{28}
\end{equation}
Matrices $\beta _\mu$, (27), realize the $6$-dimensional representation of DKP algebra. We notice that
there are $5$- and $10$-dimensional representations of DKP algebra which are used for the description of
scalar and vector particles in the framework of the first-order formalism in $(3+1)-D$. It easy to check that matrices
$\beta_1^2$, $\beta_2^2$, $\beta_3^2$ are projection operators.

Let us verify the U(1)-invariance of Eq.(25) using the method of \cite{Harish}. The gauge transformations under U(1)-group are given by
\begin{equation}
\Psi'(x)=\Psi(x)+\overline{P}\phi(x),
\label{29}
\end{equation}
where the matrix $\overline{P}$ is projection matrix ($\overline{P}^2=\overline{P}$), and reads
\begin{equation}
\overline{P}=\varepsilon^{\mu,\mu}.
\label{30}
\end{equation}
One may check with the aid of Eq.(22),(26) that the matrix $\overline{P}$ obeys the relations as follows:
\begin{equation}
\beta_\mu\overline{P}+\overline{P}\beta_\mu=\beta_\mu,~~~~P\overline{P}=0.
\label{31}
\end{equation}
Then using Eq.(25),(29),(31), we obtain
\begin{equation}
\left( \beta _\mu \partial _\mu + \mu P\right) \Psi'(x)=(1-\overline{P})\beta _\mu \partial _\mu\phi(x).
\label{32}
\end{equation}
The gauge transformations: $A_\mu(x)\rightarrow A_\mu (x)+\partial_\mu \Lambda (x)$ in the matrix notations lead to the function $\phi(x)$ in Eq.(29):
\[
\phi(x)=\left(
\begin{array}{c}
\mu \partial_\mu \Lambda(x)\\
0
\end{array}\right).
\]
One can verify that this function, $\phi(x)$, obeys the equation $(1-\overline{P})\beta_\mu\partial_\mu\phi(x)=0$, i.e. the gauge invariance holds in the first-order formalism considered.

\section{Hermitianizing matrix and $SO(2,1)$-covariance}

Under the $SO(2,1)$-group of coordinate transformations (the Lorentz group in $(2+1)$-D),
the wave function (21) becomes
\begin{equation}
\Psi ^{\prime }(x^{\prime })=T\Psi (x) , \label{33}
\end{equation}
where $6\times 6-$matrix $T$ gives the reducible tensor representation
of the $SO(2,1)$-group. The Lorentz covariance of Eq.(25) requires the validity of
equations
\begin{equation}
\beta_\mu TL_{\mu \nu }=T\beta _\nu,~~~~~~ PT=TP,  \label{34}
\end{equation}
where the Lorentz matrix $L=\{L_{\mu \nu }\}$ obeys the equation
$L_{\mu \alpha }L_{\nu \alpha }=\delta _{\mu \nu } $ (remember that we use Euclidian metric where
third components are imaginary).
At the infinitesimal Lorentz transformations the matrix $T$
is given by
\begin{equation}
T=1+\frac 12\varepsilon _{\mu \nu }J_{\mu \nu } , \label{35}
\end{equation}
where $J_{\mu \nu }$ are generators of the Lorentz group and
the infinitesimal Lorentz matrix is
\begin{equation}
L_{\mu \nu }=\delta _{\mu \nu }+\varepsilon _{\mu \nu } ,
\label{36}
\end{equation}
where $\varepsilon _{\mu \nu }=-\varepsilon _{\nu \mu }$ are three parameters of the group $SO(2,1)$.
From (34)-(36), we obtain equations as follows:
\begin{equation}
\beta _\mu J_{\alpha \nu }-J_{\alpha \nu }\beta_\mu =\delta
_{\alpha \mu }\beta _\nu -\delta _{\nu \mu }\beta _\alpha, ~~~~PJ_{\alpha \nu }=J_{\alpha \nu }P .
\label{37}
\end{equation}
The Lorentz group generators which obey Eq.(37) are given by
\begin{equation}
J_{\mu\nu}= \beta_\mu\beta_\nu-\beta_\nu\beta_\mu
=\varepsilon ^{\mu,\nu}-\varepsilon ^{\nu,\mu}+\varepsilon
^{[\lambda\mu],[\lambda\nu]}-\varepsilon
^{[\lambda\nu],[\lambda\mu]}
\label{38}
\end{equation}
or in the matrix form:
\[
J_{12}=\left(
\begin{array}{cccccc}
0&1&0&0&0&0 \\
-1&0&0&0&0&0 \\
0&0&0&0&0&0\\
0&0&0&0&0&0\\
0&0&0&0&0&-1\\
0&0&0&0&1&0
\end{array}\right),~~~~
J_{13}=\left(
\begin{array}{cccccc}
0&0&1&0&0&0 \\
0&0&0&0&0&0 \\
-1&0&0&0&0&0\\
0&0&0&0&0&-1\\
0&0&0&0&0&0\\
0&0&0&1&0&0
\end{array}\right),
\]
\vspace{-7mm}
\begin{equation} \label{39}
\end{equation}
\vspace{-7mm}
\[
J_{23}=\left(
\begin{array}{cccccc}
0&0&0&0&0&0 \\
0&0&1&0&0&0 \\
0&-1&0&0&0&0\\
0&0&0&0&-1&0\\
0&0&0&1&0&0\\
0&0&0&0&0&0
\end{array}\right).
\]
One can find with the help of Eq.(22) the commutation relations of generators (38):
\begin{equation}
\left[ J_{\mu \nu },J_{\alpha \beta}\right] =\delta _{\nu \alpha
}J_{\mu \beta}+\delta _{\mu \beta }J_{\nu \alpha}-\delta _{\nu
\beta }J_{\mu \alpha}-\delta _{\mu \alpha }J_{\nu \beta}.
\label{40}
\end{equation}
The Hermitianizing matrix $\eta$, which enters the Lorentz-invariant $ \overline{\Psi }(x)\Psi (x)=\Psi ^{+}(x)\eta \Psi(x)$ ($\Psi ^{+}(x)$ is the Hermitian conjugated wave function), satisfies the relations \cite{Gel'fand}
\begin{equation}
\eta \beta _m=-\beta _m^{+}\eta^{+} ,\hspace{0.5in}\eta \beta
_3=\beta _3^{+}\eta^{+},\label{41}
\end{equation}
where $m=1,2$. One obtains
\begin{equation}
\eta=\varepsilon^{m,m}-\varepsilon^{3,3}+\varepsilon^{[m3],[m3]}-
\frac{1}{2}\varepsilon^{[mn],[mn]},
\label{42}
\end{equation}
or in the form
\begin{equation}
\eta=\left(
\begin{array}{cccccc}
1&0&0&0&0&0 \\
0&1&0&0&0&0 \\
0&0&-1&0&0&0\\
0&0&0&-1&0&0\\
0&0&0&0&1&0\\
0&0&0&0&0&1
\end{array}\right),
\label{43}
\end{equation}
so that $\eta^{+}=\eta$, i.e. it is the Hermitian matrix. One may verify that $\eta$
commutes with the projection operator $P$ (see Eq.(27)), $\eta P=P\eta$.
We find the ``conjugated" wave function
\begin{equation}
\overline{\Psi }(x)=\Psi ^{+}(x)\eta=
\left(\mu A_\nu,-F_{\mu\nu}\right). \label{44}
\end{equation}
From Eq.(25) one obtains the ``conjugated" equation
\begin{equation}
\overline{\Psi }(x)\left( \beta _\mu \overleftarrow{\partial} _\mu
-\mu P^+\right) =0 .
\label{45}
\end{equation}
One may check the relationship
\begin{equation}
\overline{\Psi }(x)P^+\Psi (x)=\overline{\Psi }(x)P\Psi (x), \label{46}
\end{equation}
but $P^+\neq P$. Therefore, from Eq.(25),(46), we obtain the conservation of
the charge: $\partial_\mu\left(\overline{\Psi }(x)\beta_\mu\Psi (x)\right)=0$.
For our case of neutral fields (($\textbf{A},A_0$) are real values), we have from (21), (27) and (44):
$\overline{\Psi }(x)\beta_\mu\Psi (x)=0$.

\section{Solutions to RWE}

Let us find solutions to Eq.(25) corresponding to definite energy and
momentum. Thus, in the momentum space Eq.(25) takes the form
\begin{equation}
\left(i\hat{p}+\mu P \right)\Psi(p)=0 ,
\label{47}
\end{equation}
where $\hat{p}=\beta_\mu p_\mu$, $p_\mu$ is a energy-momentum, $p_\mu=(p_1,p_2,ip_0)$, $p_\mu^2=-\mu^2$.
The matrix of the equation (47)
\begin{equation}
\Lambda= i\hat{p}+\mu P
\label{48}
\end{equation}
reads
\begin{equation}
\Lambda=\left(
\begin{array}{cccccc}
0&0&0&ip_2&ip_3&-i\mu \\
0&0&0&-ip_1&i\mu&ip_3 \\
0&0&0&-i\mu&-ip_1&-ip_2\\
ip_2&-ip_1&0&\mu&0&0\\
ip_3&0&-ip_1&0&\mu&0\\
0&ip_3&-ip_2&0&0&\mu
\end{array}\right).
\label{49}
\end{equation}
One can verify that the $6\times 6$-matrix $\Lambda$ (49) obeys the `minimal' polynomial equation
\begin{equation}
\Lambda\left(\Lambda^2-\mu^2\right)\left(\Lambda-2\mu\right)=0.
\label{50}
\end{equation}
It is obvious that the matrix
\begin{equation}
\Pi=N\left(\Lambda^2-\mu^2\right)\left(\Lambda-2\mu\right),
\label{51}
\end{equation}
where $N$ is a constant, satisfies the equation $\Lambda\Pi=0$,
so that the every column of the matrix $\Pi$ is the
solution to Eq.(25). It is convenient to require that $\Pi$ is the
projection operator \cite {Fedorov}. From the relation $\Pi^2=\Pi$, we obtain the normalization
constant $N=1/(2\mu^3)$. The projection operator (51) gives solutions to Eq.(25) for definite energy-momentum.

Now, we define pseudovector generators of $SO(2,1)$-group as follows \cite{Dunne}:
\begin{equation}
J_\mu=\frac{1}{2}\varepsilon_{\mu\nu\alpha}J_{\nu\alpha},
\label{52}
\end{equation}
where $J_{\nu\alpha}$ are given by (39). It is easy to check that generators (52) obey the
commutation relations:
\begin{equation}
\left[J_\mu,J_\nu\right]=\varepsilon_{\nu\mu\alpha}J_{\alpha},~~~~
\left[J_\mu,\hat{p}\right]=\varepsilon_{\mu\nu\alpha}p_{\nu}\beta_\alpha .
\label{53}
\end{equation}
The Pauli-Lubanski pseudoscalar is defined by (see \cite{Dunne})
\begin{equation}
W=p_\mu J_\mu,
\label{54}
\end{equation}
and reads
\begin{equation}
W=\left(
\begin{array}{cccccc}
0&-ip_3&ip_2&0&0&0\\
ip_3&0&-ip_1&0&0&0 \\
-ip_2&ip_1&0&0&0&0\\
0&0&0&0&ip_1&-ip_2\\
0&0&0&-ip_1&0&ip_3\\
0&0&0&ip_2&-ip_3&0
\end{array}\right).
\label{55}
\end{equation}
The $W$ ia a Hermitian matrix. One can verify that $W$ commutates with operators $\widehat{p}$, $P$ and $\Lambda$: $[W,\widehat{p}]=0$, $[W,P]=0$, $[W,\Lambda]=0$.
Operator (55) obeys the `minimal' polynomial equation:
\begin{equation}
W\left( W^2-p^2\right) =0 ,
\label{56}
\end{equation}
and we use the notation $p^2\equiv p^2_\mu$. According to the general method \cite{Fedorov}, we find
the projection operators extracting physical states with spins $\pm 1$ and non-physical states with spin $0$:
\begin{equation}
S_{(\pm 1)}=\frac 12\Sigma\left( \Sigma\pm 1\right)
,\hspace{0.5in}S_{(0)}=1-\Sigma^2,
\label{57}
\end{equation}
where
\[
\Sigma=\frac{W}{\sqrt{p^2}}=-i\frac{W}{\mu}.
\]
Projection operators (57) obey the relations: $S_{(\pm 1)}^2=S_{(\pm 1)}$, $S _{(\pm
1)}S_{(0)}=0$, $S_{(0)}^2=S_{(0)}$, $[\Pi,S_{(\pm 1)}]$=0, $[\Pi,S_{(0)}]$=0, $S_{(+1)}+S_{(- 1)}+S_{(0)}=1$.
From Eq.(55)(57), we obtain
\begin{equation}
S_{(\pm 1)}=\frac{1}{2\mu^2}\left(
\begin{array}{cccccc}
-p_2^2-p_3^2&p_1p_2\mp\mu p_3&p_1p_3\pm\mu p_2&0&0&0\\
p_1p_2\pm\mu p_3&-p_1^2-p_3^2&p_2p_3\mp\mu p_1&0&0&0 \\
p_1p_3\mp\mu p_2&p_2p_3\pm\mu p_1&-p_1^2-p_2^2&0&0&0\\
0&0&0&-p_1^2-p_2^2&p_2p_3\pm\mu p_1&p_1p_3\mp\mu p_2\\
0&0&0&p_2p_3\mp\mu p_1&-p_1^2-p_3^2&p_1p_2\pm\mu p_3\\
0&0&0&p_1p_3\pm\mu p_2&p_1p_2\mp\mu p_3&-p_2^2-p_3^2
\end{array}\right).
\label{58}
\end{equation}
It follows from Eq.(58) that spin operators $S_{(+ 1)}$ and $S_{(- 1)}$ differ only by the sign in the mass $\mu$ according to Poincar\'{e} algebra representation \cite{Deser}, \cite{Dunne}.
One finds projection operators
\begin{equation}
\Delta_{\pm 1} =\Pi S_{(\pm 1)} ,~~~~\Delta _0=\Pi S_{(0)}
\label{59}
\end{equation}
extracting solutions to Eq.(25) for definite energy-momentum and spin. The projection operator $\Delta_{\pm 1}$ extracts physical solutions and also defines the density matrix which can be used for covariant quantum calculations \cite{Fedorov}.

\section{Schr\"{o}dinger form of equations}

To obtain the Schr\"{o}dinger form of equations and quantum-mechanical Hamiltonian,
we exclude the non-dynamical components from the wave function (21).
For this purpose, we represent Eq.(2) and equation $F_{\mu\nu}=\partial_\mu A_\nu-\partial_\nu A_\mu$ as follows:
\[
i\partial_t A_1 =F_{13}-\partial_1 A_3,~~~~~~i\partial_t A_2=F_{23}-\partial_2 A_3,
\]
\vspace{-8mm}
\begin{equation}
\label{60}
\end{equation}
\vspace{-8mm}
\[
i\partial_tF_{31}=-\partial_2F_{12}+i\mu F_{23},~~~~i\partial_tF_{23}=-\partial_1F_{12}-i\mu F_{31},
\]
\begin{equation}
F_{12}=\partial_1A_2-\partial_2A_1.
\label{61}
\end{equation}
Eq.(60) show the evolution of the dynamical components in time, but Eq.(61) does not contain derivative in time, and therefore, $F_{12}$ is non-dynamical component of fields. In addition, there is not evolution in time of the component of the potential $A_3$ ($A_3=iA_0$). It is possible to use the gauge condition $A_3=0$ \cite{Dunne} but it violates the Lorentz-covariance. Instead of this gauge, we use the Lorentz condition $\partial_\mu A_\mu=0$ which defines the evolution in time of the $A_3$-component:
 \begin{equation}
i\partial_t A_3 =\partial_1 A_1+\partial_2 A_2.
\label{62}
\end{equation}
One can exclude the auxiliary (non-dynamical) component $F_{12}$ from Eq.(60).
Replacing the non-dynamical components $F_{12}$ from Eq.(61) into Eq.(60), and taking into account Eq.(62), we obtain
\begin{equation}
i\partial _t \left(
\begin{array}{c}
 \mu A_1\\
\mu A_2\\
\mu A_3\\
F_{31}\\
F_{23}
\end{array}\right)=
\left(
\begin{array}{c}
 -\mu\left(F_{31}+\partial_1 A_3\right)\\
\mu\left(F_{23}-\partial_2 A_3\right)\\
\mu\left(\partial_1 A_1+\partial_2 A_2\right)\\
\frac{i}{\mu}\left(\partial_1\partial_2F_{31}-\partial_2^2F_{23}\right)+i\mu F_{23}\\
\frac{i}{\mu}\left(\partial_1^2F_{31}-\partial_1\partial_2F_{23}\right)-i\mu F_{31}
\end{array}\right).
 \label{63}
\end{equation}
Introducing 5-component wave function
\begin{equation}
\Phi (x)=\left(
\begin{array}{c}
\mu A_\nu (x)\\
F_{31}(x)\\
F_{23}(x)
\end{array}
\right),
\label{64}
\end{equation}
Eq.(63), with the help of the elements of the entire matrix algebra Eq.(22), can be
written in the Schr\"{o}dinger form
\begin{equation}
i\partial_t\Phi(x)={\cal H}\Phi(x),
\label{65}
\end{equation}
where the Hamiltonian is given by
\[
{\cal H}=\mu\left[\varepsilon^{n,[n3]}+i\left(\varepsilon^{[31],[23]}-
\varepsilon^{[23],[31]} \right)\right]+\partial_1\left(\varepsilon^{3,1}-
\varepsilon^{1,3}\right)+\partial_2\left(\varepsilon^{3,2}-
\varepsilon^{2,3}\right)
\]
\vspace{-8mm}
\begin{equation} \label{66}
\end{equation}
\vspace{-8mm}
\[
+\frac{i}{\mu}\left[\left(\varepsilon^{[31],[31]}-
\varepsilon^{[32],[32]}\right)\partial_1\partial_2-\varepsilon^{[31],[23]}\partial_2^2
+\varepsilon^{[23],[31]}\partial_1^2
\right],
\]
where we imply the summation on index $n$.
The wave function (64) contains only dynamical components of gauge fields, and therefore the Schr\"{o}dinger picture has some advantages by considering field interactions.
To obtain solutions to Eq.(65) for free fields, we consider the momentum space, where the Hamiltonian (66) becomes
\begin{equation}
{\cal H}=\left(\begin{array}{ccccc}
0&0&-ip_1&-\mu&0\\
0&0&-ip_2&0&\mu \\
ip_1&ip_2&0&0&0\\
0&0&0&-\frac{i}{\mu}p_1p_2&\frac{i}{\mu}\left(p_2^2+\mu^2\right)\\
0&0&0&-\frac{i}{\mu}\left(p_1^2+\mu^2\right)&\frac{i}{\mu}p_1p_2
\end{array}\right).\label{67}
\end{equation}
One may verify that the matrix (67) satisfies the `minimal' polynomial equation
\begin{equation}
{\cal H}^2\left({\cal H}^2-p_n^2\right)\left[{\cal H}^2-\left(p_n^2+\mu^2\right)\right]=0,
\label{68}
\end{equation}
and $p_n^2=p_1^2+p_2^2$. It follows from Eq.(68) that the Hamiltonian ${\cal H}$ possesses the eigenvalues:
$0$, $\pm\sqrt{p_n^2}$, $\pm \sqrt{p_n^2+\mu^2}$.
From equation (68), we obtain projection operators extracting physical states with positive
and negative energies ($p_0=\pm \sqrt{p_n^2+\mu^2}$):
\begin{equation}
\Sigma_{\pm}=\frac{\left({\cal H}\pm
\sqrt{p_n^2+\mu^2}\right){\cal H}^2\left({\cal H}^2-p_n^2\right)}{2\mu^2\left(p_n^2+\mu^2\right)^{3/2}},
\label{69}
\end{equation}
so that $\left(\Sigma_{\pm}\right)^2=\Sigma_{\pm}$, and
\begin{equation}
{\cal H}\Sigma_{\pm}=\pm \sqrt{p_n^2+\mu^2}\Sigma_{\pm}.
\label{70}
\end{equation}
The projection operator (69) can be used for constructing physical states in the space of wave functions
(64). Indeed, the wave function $\Phi_{\pm}=\Sigma_{\pm}\Phi_0$, where  $\Phi_0$ is a constant non-zero 5-column,
is an eigenfunction of the Hamiltonian ${\cal H}$ with physical eigenvalues $\pm \sqrt{p_n^2+\mu^2}$:
\begin{equation}
{\cal H}\Phi_{\pm}=\pm \sqrt{p_n^2+\mu^2}\Phi_{\pm}.
\label{71}
\end{equation}
Thus, the wave function $\Phi_{\pm}$ defines physical states in the Schr\"{o}dinger picture.

\section{Conclusion}

We have obtained the canonical and Belinfante energy momentum tensors and non-conserved dilatation current which is proportional to the topological mass of the gauge field. It was demonstrated that the `scale dimensionality' of gauge fields is $d=1/2$ in $(2+1)$-D. This is different from fields in $(3+1)$-D.
The field equations were represented in the form of the first-order RWE, and we obtained $6\times 6$-matrices of RWE which obey DKP algebra. Solutions to RWE for a free fields with definite energy-momentum and spin were obtained in the form of projection matrices. These solutions can be used for different covariant quantum-mechanical calculations.
The Schr\"{o}dinger form of equations and the quantum-mechanical Hamiltonian obtained can be explored in the first-order formalism. Thus, projection operators extracting physical states of free fields can be used in 5-dimensional space of dynamical components of gauge fields.
It should be noted that DKP form of equations was also considered in non-Abelian theory (QCD) \cite{Gribov}.

\vspace{7mm}
\textbf{\textbf{Acknowledgments}}
\vspace{7mm}

I wish to thank Prof. S. Deser for correspondence.

\vspace{7mm}
\textbf{Appendix A}
\vspace{7mm}

To demonstrate that the gauge field possesses the mass $\mu$, we represent equation of motion (2) in the momentum space.
Replacing the strength tensor $F_{\mu\nu}=\partial_\mu A_\nu-\partial_\nu A_\mu$ in Eq.(2), one obtains the equation for vector-potential in the momentum space:
\[
\left[i\left(p^2\delta_{\mu\nu}-p_\mu p_\nu\right)+\mu\varepsilon_{\mu\alpha\nu}p_\alpha\right]A_\nu=0,~~~~~~~~~~~~~~~~~~~~~~~~~~~~~(A.1)
\]
where $p^2=p_1^2+p_2^2-p_0^2$. Introducing the antisymmetric tensor $\epsilon_{\mu\nu\alpha}=i\varepsilon_{\mu\nu\alpha}$ ($\epsilon_{123}=1$), we rewrite Eq.(A.1) in the form
\[
M_{\mu\nu}A_\nu=0,~~~~~~~~~~~~~~~~~~~~~~~~~~~~~~~~~~~~~~~~~~~~~~~~~~~~~~~~~~~~~~~(A.2)
\]
where
\[
M_{\mu\nu}=p^2\delta_{\mu\nu}-p_\mu p_\nu-\mu\epsilon_{\mu\alpha\nu}p_\alpha.~~~~~~~~~~~~~~~~~~~~~~~~~~~~~~~~~~~~~~~(A.3)
\]
Equation (A.2) possesses nontrivial solutions if determinant of the matrix $M=\left\{M_{\mu\nu}\right\}$ equals zero.
We find the eigenvalues of the matrix $M$ by solving the characteristic equation $\det\left(M-\lambda\right)=0$:
\[
\det\left[M-\lambda\right)=-\lambda\left(\lambda^2-2\lambda p^2+p^2\left(p^2+\mu^2\right)\right]=0.~~~~~~~~~~~~~~(A.4)
\]
From Eq.(A.4), we obtain eigenvalues of the matrix $M$:
\[
\lambda_1=0,~~~~\lambda_2=p^2+\sqrt{-p^2\mu^2},~~~~\lambda_3=p^2-\sqrt{-p^2\mu^2}.~~~~~~~~~~~~(A.5)
\]
The requirement $\lambda_2=0$ leads to the dispersion relation $p^2=-\mu^2$ defining the mass $\mu$ of the gauge field.
We notice that the matrix $M$ obeys the equation as follows:
\[
M\left[M^2-2p^2 M +p^2\left(p^2+\mu^2\right)\right]=0,~~~~~~~~~~~~~~~~~~~~~~~~~~~~~~~~(A.6)
\]
in accordance with the Hamilton$-$Cayley theorem.

\vspace{7mm}
\textbf{Appendix B}
\vspace{7mm}

In order to obtain Schwinger's \cite{Schwinger} symmetrical energy-momentum tensor, we rewrite Lagrangian (1) in the
curve coordinate system:
\[
{\cal L} =-\frac{1}{4}g^{\mu\alpha}g^{\beta\sigma}F_{\alpha\beta}F_{\mu\sigma}+
\frac{\mu}{4}\varepsilon_{\rho\sigma\beta}g^{\rho\mu}g^{\sigma\nu}g^{\beta\alpha}F_{\mu\nu}A_\alpha.~~~~~~~~~~~~~~~~~~(B.1)
\]
Then the symmetrical energy-momentum tensor obtained from varying action by the metric tensor $T^{sym}_{\mu\nu}=\delta S/\delta g^{\mu\nu}$ can be find from the equation
\[
T^{sym}_{\mu\nu}=2\left[\frac{\partial {\cal L}}{\partial g^{\mu\nu}}-
\partial_\alpha \frac{\partial {\cal L}}{\partial(\partial_\alpha g^{\mu\nu})}\right]-g_{\mu\nu}{\cal L}.~~~~~~~~~~~~~~~~~~~~~~~~~~~(B.2)
\]
Replacing Eq.(B.1) into (B.2), we obtain
\[
T_{\mu\nu}^{sym}=-F_\mu^{~\alpha}F_{\nu\alpha}+\frac{\mu}{4}\varepsilon_\mu^{~\alpha\beta}\left(
2F_{\nu\alpha}A_\beta+F_{\alpha\beta}A_\nu\right)-g_{\mu\nu}{\cal L}.~~~~~~~~~~~(B.3)
\]
Returning to Euclidian metric, one can verify that according to \cite{Deser} the symmetrical Chern$-$Simons energy-momentum tensor vanishes:
\[
T_{\mu\nu}^{C-S}=\frac{\mu}{4}\varepsilon_{\mu\alpha\beta}\left(
2F_{\nu\alpha}A_\beta+F_{\alpha\beta}A_\nu\right)-\delta_{\mu\nu}\left( \frac{\mu}{4}\varepsilon_{\rho\sigma\alpha}F_{\rho\sigma}A_\alpha\right)=0.~~~(B.4)
\]
Equation (B.4) is valid because of the identity:
\[
\varepsilon_{\mu\alpha\beta}\left(
2F_{\nu\alpha}A_\beta+F_{\alpha\beta}A_\nu\right)=
\delta_{\mu\nu} \varepsilon_{\rho\sigma\alpha}F_{\rho\sigma}A_\alpha,~~~~~~~~~~~~~~~~~~~~~~~~~(B.5)
\]
which can be checked by evaluating the components in left and right sides of the equation.

\end{document}